\documentclass[%
preprint,
 amsmath,amssymb,
aps,
]{revtex4-2}
\usepackage{float}
\usepackage{graphicx}
\usepackage{dcolumn}
\usepackage{bm}
\usepackage{subfigure}
\usepackage{diagbox}
\usepackage{url}

\begin{document}

\preprint{APS/123-QED}

\title{Higher-order topological states in photonic Thue-Morse quasicrystals: quadrupole insulator and a new origin of corner states}

\author{Langlang Xiong$^{1,3}$}

\author{Yu Zhang$^{2,3}$}%

\author{Yufu Liu$^{2,3}$}%

\author{Yaoxian Zheng$^{4}$}%

\author{Xunya Jiang$^{1,2,3}$}%
 \email{jiangxunya@fudan.edu.cn}
 \affiliation{
	$^{1}$Institute of Future Lighting, Academy for engineering and technology, Fudan University, Shanghai, 200433, China
}%
\affiliation{
	$^{2}$Department of Illuminating Engineering and Light Sources, School of Information Science and Engineering, Fudan University, Shanghai, 200433, China
}%
\affiliation{
	$^{3}$Engineering Research Center of Advanced Lighting Technology, Fudan University, Ministry of Education, Shanghai, 200433, China
}%

\affiliation{
	$^{4}$THz Technical Research Center and College of Physics and Optoelectronic Engineering, Shenzhen University, Shenzhen, 518060, China
}%

\date{\today}

\begin{abstract}
Corner states (CSs) in higher-order topological insulators (HOTIs) have recently been of great interest in both crystals and quasicrystals. In contrast to electronic systems, HOTIs have not been found in photonic quasicrystals (PQCs). Here, we systemically study the higher-order topology in the two-dimensional Thue-Morse photonic quasicrystals (TM-PQCs). Not only the topological phase transition and the non-trivial CSs with fractional charge induced by multipole moments, but also a new type of CSs are found due to the complex structure of TM-PQCs near corners. The different origins of these CSs are also analyzed based on the tight-binding model. Our work opens the door to explore richer HOT physics beyond photonic crystals and the robustness of CSs in PQC shows the potential for applications.

\end{abstract}

\maketitle



\textit{Introduction.}---In the research of topological systems\cite{qi2011topological,hasan2010colloquium, bansil2016colloquium,chiu2016classification,ozawa2019topological,rider2019perspective,lu2014topological,khanikaev2017two,wu2017applications,hu2020topological,xiong2021resonance},
higher-order topology (HOT) has become a new hot-spot since it could lead to unique topological states beyond traditional bulk-boundary correspondence \cite{kim2020recent,xie2021higher,liu2021bulk,xie2018second,xie2019visualization,xiong2022topological}. Specifically, a kind of two dimensional (2D) higher-order topological insulators (HOTIs), i.e., quadrupole insulators (QIs) \cite{benalcazar2017quantized,he2020quadrupole,benalcazar2017electric}, whose topological invariant, quadrupole moments $q_{xy}$, is quantized to 0 or 0.5 if the system presents fourfold rotation symmetry $C_4$ or mirror symmetries $M_x := x \to -x$ and $M_y := y \to -y$, and non-zero $q_{xy}$ can give rise to the zero dimensional (0D) non-trivial corner states (CSs), namely type-I CSs.
Besides, additional type-II CSs that cause by long-range interactions also have been found \cite{li2020higher,shen2021investigation,xu2020general}.

Very recently, the concept of HOTIs has extended from periodic crystals to quasicrystals (QCs) and aperiodic crystals \cite{chen2020higher,lv2021realization,hu2021disorder,wang2021strStructural,huang2021generic,Tobias2022Fractal}, which also show non-trivial zero-energy CSs in the 2D quantum system.
However, in contrast to the study of HOTIs in electronic systems based on the tight-binding model (TBM), the realizations and physical properties of HOTIs in photonic quasicrystals (PQCs) have not been studied, which have abundant applications in reality. Even more, the research for HOT CSs of the electronic systems concentrates on type-I CSs, so we still cannot answer such questions, such as ``in QCs, can we realize richer CSs, or even find different origins of CSs beyond the two types of CSs in crystals?".

In this work, we systemically investigate the HOT properties of 2D Thue-Morse (TM) PQCs \cite{zhang2021fractal, luigi2007two}. We first construct a TM-PQCs with two kinds of dielectric rods, then a TBM bases on the TM-PQC is built. By tuning difference parameters of TBM, the HOT phase transition with non-zero $q_{xy}$ is found. Moreover, the different origins of CSs in PQCs are revealed by using weak-coupling limit \cite{xiong2022topological,liu2017novel,li2020higher}. From the different origins, we demonstrate that, besides the type-I CSs with fractional charges and the type-II CSs from long-range interactions, there is a new type of CSs from the complex structure of TM-PQCs which is independent of long-range interactions and beyond the periodic systems. Finally, based on the strict numerical simulation, we show that all these CSs can exist in real TM-PQCs and CSs from TM-PQCs could be more robust compared with those in photonic crystals (PhCs). This work is valuable in expanding the understanding of HOT phases beyond periodic photonic systems and observing CSs in PQCs with special properties which can be further utilized to design the aimed devices with topological protection.

\begin{figure*}[htbp] \centering \includegraphics[width=1\textwidth]{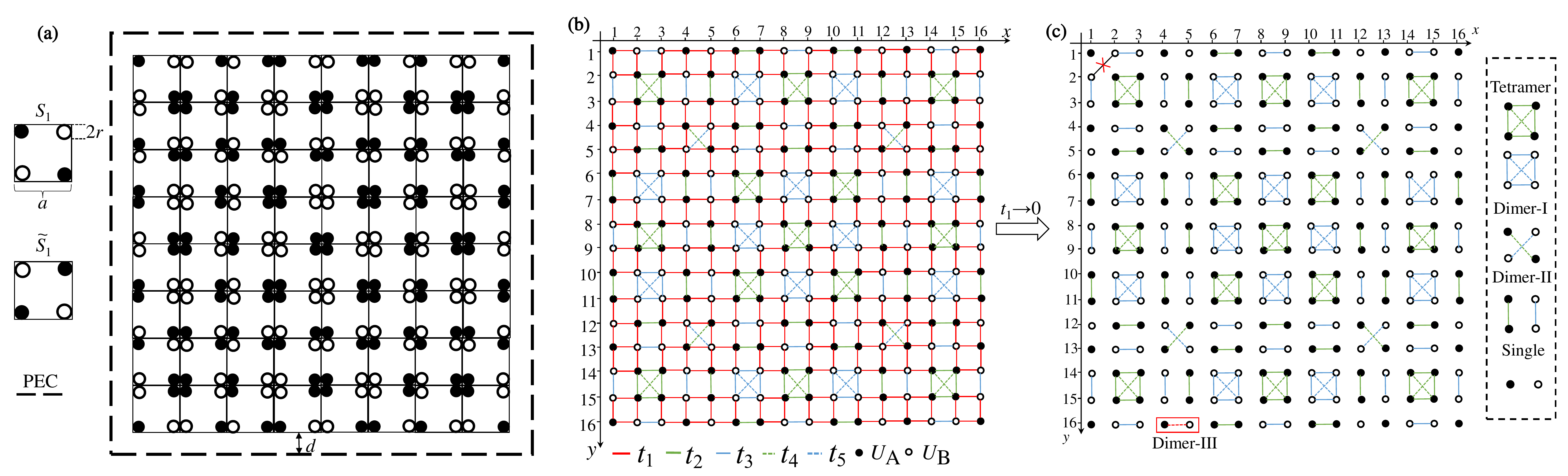} \caption{
		(a) The $S_4$ TM-PQCs with rods-A and rods-B, the PEC boundary is marked with black dash lines; (b) The TBM bases on the 2D TM-PQC; (c) The limiting case of TBM with $t_1=0$.
}\label{fig1} \end{figure*}

\textit{Model and topological phase transition.}---
A 2D TM sequence $S_N$ of order $N$ can be deduced by:
\begin{equation}
	S_{N}=\left[\begin{array}{ll}
		S_{N-1} & \tilde{S}_{N-1} \\
		\tilde{S}_{N-1} & S_{N-1}
		\end{array}\right] \quad \text { and } \quad  S_1 = \left[\begin{array}{ll} \rm{A\ B} \\ \rm{B\ A} \end{array}\right],\label{equ1}
\end{equation}
where $\tilde{S}_{N-1}$ is obtained by exchanging A and B in $S_{N-1}$. The photonic model of 2D square TM sequence can be generated by placing different dielectric rods in the square lattice. In particular, TM photonic structures of basic sequences $S_1$ and $\tilde{S}_{1}$ is shown in Fig. \ref{fig1}(a), and we mark them with basic square cells. The four corners of a square cell contain four dielectric rods in the air with radius $r=0.12a$, where $a$ is the side-length of square cell. The dielectric rods are divided into two types: rods-A with relative permittivities $\varepsilon_A$ and relative permeability $\mu_A$, and rods-B with $\varepsilon_B$ and $\mu_B$. Then, we can use Eq. (\ref{equ1}) and square cell $S_1$ to deduce higher-order TM-PQC, e.g., Fig. \ref{fig1}(a) shows a PQC of $S_4$ TM sequence. We can find a TM-PQC of even order has $C_4$ and $M_{x(y)}$ symmetries, whereas TM-PQC of odd order does not.

In recent works \cite{xie2018second,xiong2022topological,li2020higher}, it's found that TBM is a good platform to reveal the origins of topological states in the photonic systems for the lower bands. Following this path, we also construct a TBM bases on TM-PQC which is shown in Fig. \ref{fig1}(b). The on-site energy of rods-A(B) is $U_{A(B)}$. Considering the frequency difference of Mie resonances of two kinds of rods, the nearest-neighbor coupling between rod-A and rod-B $t_1$ is supposed to be a small value generally.
The inter-cell coupling between two rods-A(or B) is $t_{2(3)}$, and the next-nearest-neighbor (NNN) coupling between two rods-A (or B) of inter-cell on the diagonal direction $t_{4(5)}$ which can open the gap at energy $E=0$.
In Fig. \ref{fig1}(c), for the convenience of our further study, TBM of the limiting case with $t_1=0$ is shown, in which the entire structure splits into third types of isolated clusters: two tetramers, four dimers, and two singles. The dimers in diagonal directions and in horizontal/vertical directions are marked with dimers-I and dimers-II, respectively. Such split model is helpful for us to analyze the origins of different topological states in the next section. Note that the coupling between the same type of rods inside one cell is neglected since the large distance between them and we emphasize this by a red cross in the left-up corner in Fig. \ref{fig1}(c).

\begin{figure*}[htbp] \centering \includegraphics[width=1\textwidth]{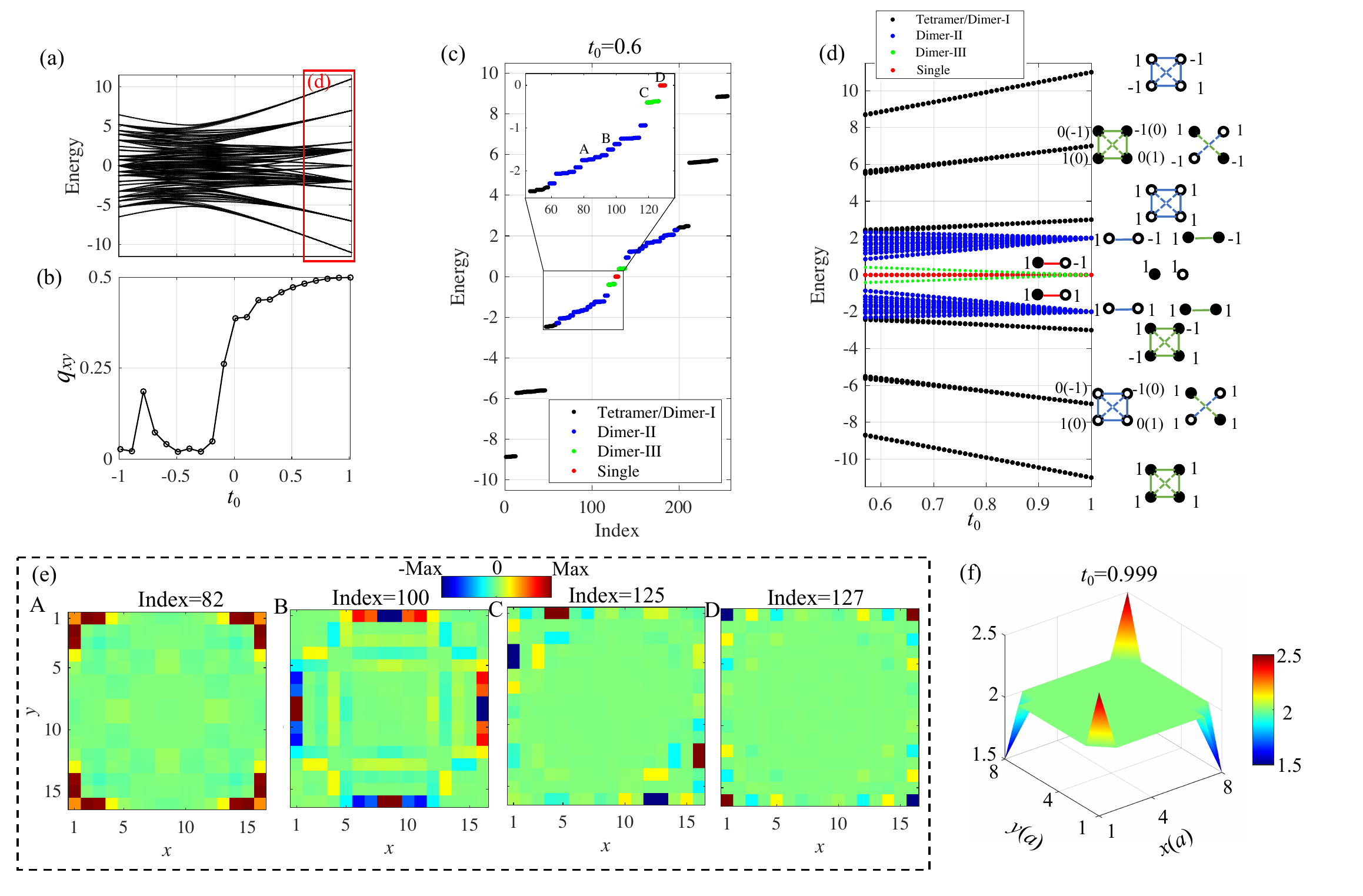} \caption{
		(a) The band structure of $S_4$ TM-QC with open boundary condition versus $t_0 \in [-1,1]$, other parameters are set as $t_1=1-t_0$, $t_2=t_3=1+t_0$, $t_4=-t_5=3.5t_1$, and $U_A=U_B=0$; (b)  The quadrupole moment $q_{xy}$ versus different $t_0$; (c) The band structure of $S_4$ TM-QC with $t_0=0.6$, four typical localized states are marked with A-D; (d) The local enlarged (a) for the topological non-trivial region; (e) Four localized states that marked in (c); (f) The distributions of LDOS with $t_0=0.999$.
}\label{fig2} \end{figure*}

To focus on the process of topological phase transition, we introduce a variable $t_0$ and set $t_1=1-t_0$, $t_2=t_3=1+t_0$, $t_4=-t_5=3.5t_1$, and $U_A=U_B=0$. In Fig. \ref{fig2}(a), the band structure of open boundary condition versus $t_0 \in [-1,1]$ are drawn. By using real space method \cite{resta1998quantum,wheeler2019many,he2020quadrupole, RSM,kang2019many, SI}, we calculate the quadrupole moment $q_{xy}$ of the gap at $E=0$ versus different $t_0$ in Fig. \ref{fig2}(b).
We can see $q_{xy}$ jumps from $0$ to $0.5$ for $t_0$ from $-1$ to $1$, which means a topological phase transition, and the main jumping is happened near $t_0=0$, i.e., $t_1=t_2=t_3$.
Specifically, we choose $t_0=0.6$ and its band structure is shown in Fig. \ref{fig2}(c), where type-I CSs at zero energy that are protected by non-zero quadrupole moment are marked with red dots. The topological non-triviality of CSs is also confirmed by the index requirement from the filling anomaly theory \cite{he2020quadrupole} that the indices of CSs in our model are $127-130$. Fig. \ref{fig2}(e)-D shows a typical distribution of type-I CS.
What's more, we calculate the sum of the lowest half states to show direct distributions of fractional corner charges \cite{SI}, which can be proved to be equal to quadrupole moment \cite{benalcazar2017quantized}.
In Fig. \ref{fig2}(f), we set a large $t_0=0.999$ to obtain more localized CSs, and the distributions of corner charges are shown.
It's found that the fractional corner charge is quantized to 0.5 which is because of $C_4$ and $M_{x(y)}$ symmetries, and the edge charge keep zero because of $C_2$ symmetry \cite{he2020quadrupole}.

\textit{The new types of CSs in TM-QC.}---
Besides type-I CSs that are protected by quadrupole moment, some other localized states in TM-QC also can be observed. For example, we choose four typical localized states that are marked in Fig. \ref{fig2}(c) as A-D, and the four states are drawn in Fig. \ref{fig2}(e). In addition to non-trivial type-I CS has been mentioned in Fig. \ref{fig2}(e)-D, there are two CSs in Fig. \ref{fig2}(e)-A and (e)-C and one edge state (ES) in Fig. \ref{fig2}(e)-B.
The two CSs in Fig. \ref{fig2}(e)-A and (e)-C are not very localized as type-I CSs and are more likely to be recognized as the type-II CSs \cite{li2020higher, xiong2022topological}. However, in our model the NNN coupling between the same rods in one cell is neglected which is essential for the existence of type-II CSs, so the physical origin of CSs in Fig. \ref{fig2}(e)-A and (e)-C should be reconsidered carefully.

First, we need to go back to Fig. \ref{fig1}(c) in the limiting case $t_1=0$ to reveal the different origins of the new type of CSs in \ref{fig2}(e)-A and (e)-C. According to Fig. \ref{fig1}(c), we can obtain some basic clues of the state origin from the local split structures. Second, we can analyze the origin of states in Fig.\ref{fig2} more carefully.
Specifically, we zoom in the area of the red rectangle in Fig. \ref{fig2}(a) and show the area in Fig. \ref{fig2}(d), where $t_1=1-t_0$ is a small value comparing with other coupling terms since $t_0>0.6$.  In Fig. \ref{fig2}(d), we mark the states with different colors according to their different cluster origins, e.g. the states from tetramers/dimers-I, dimers-II, dimers-III, and singles are marked with black, blue, green, and red dots, respectively. On right side of Fig. \ref{fig2}(d) near every band, in the limit $t_0=1$ ($t_1=0$), we also show the symmetry property of the states for different clusters. Next, we will introduce more details of these states from different clusters, which could be solved theoretically.
First, the tetramers support six eigenstates, two singlet quadrupole mode with $E=-t_{4(5)}+2t_{2(3)}$, two doublet of dipolar modes with $E=t_{4(5)}$, and two singlet monopolar mode with $E=-t_{4(5)}-2t_{2(3)}$. Second, the dimers-I and dimers-II support eight dipolar modes: symmetric modes with $E=-t_{2,3,4,5}$ and antisymmetric modes with $E=t_{2,3,4,5}$. Third, the singles support two eigenstates with $E=0$ since $U_A=U_B=0$.

If we introduce non-zero but small $t_1$, we can find topological phenomena in our 2D-TM systems, like or unlike 2D crystals.
For example, when we set $t_0=0.6$, the ESs shown in Fig. \ref{fig2}(e)-B  and the type-I CSs shown in Fig. \ref{fig2}(e)-D could be observed, similar to the crystals. From the field distribution and the energy, we find that the ESs are from dimers-II structure and located at the system edges. For the type-I CSs, we find they are from singles at the corners of the system. So the origin of these ESs and CSs are similar to crystals. However, since the structure of 2D-TM systems are much more complex than 2D crystals, some new states could be observed in the gaps, like the states shown in Fig. \ref{fig2}(e)-A and Fig. \ref{fig2}(e)-C which need to be carefully investigated and are beyond 2D crystals.

\begin{figure*}[htbp] \centering \includegraphics[width=1\textwidth]{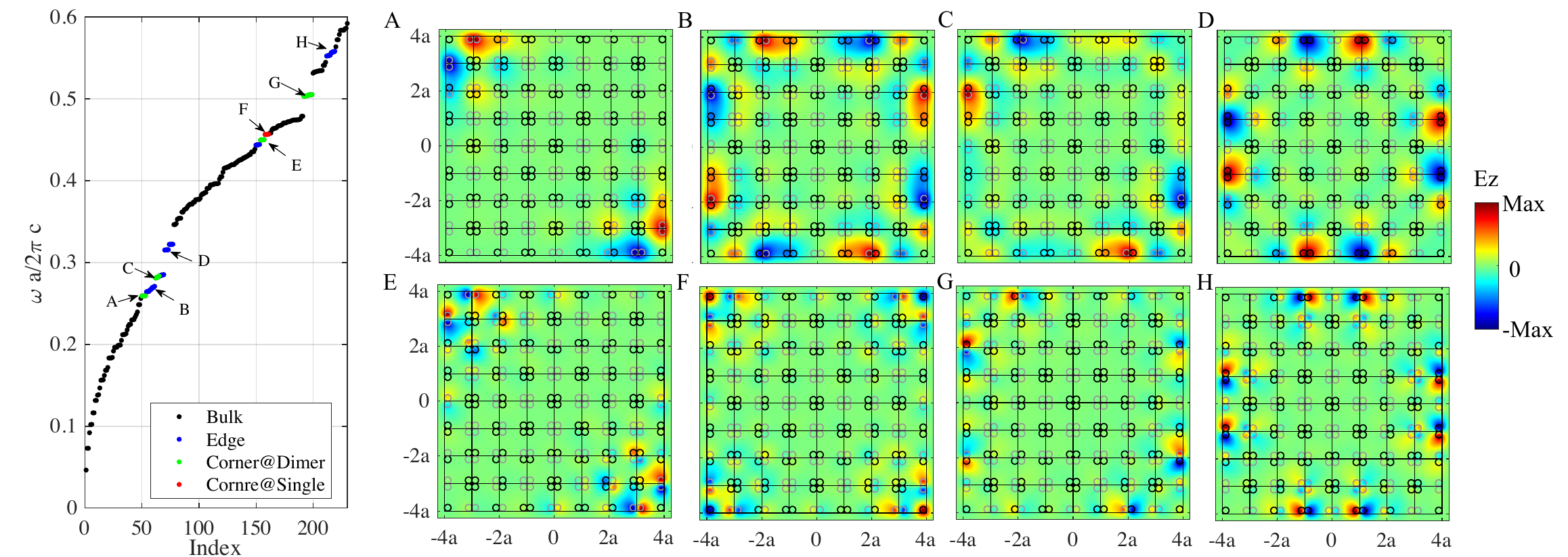} \caption{
	Band structure of a $S_4$ TM-PQC with $\varepsilon_A=10$, $\varepsilon_B=16$, and PEC boundary. The distance between PEC and PQC is $d=0.25a$. Eight typical states are marked with A-H, and the $E_z$ field distributions of those states are also shown on the right, where black (grey) circles are rods-A(B).
}\label{fig3} \end{figure*}

For the state in Fig. \ref{fig2}(e)-A, we need go back to Fig. \ref{fig1}(c). From the field distribution of the state and the structure on the corner in Fig. \ref{fig1}(c), we can see that the state originated from the coupling of two dimers-II near the corners by the corner single. We note that this type of CSs is different from the type-II CSs in crystals \cite{xiong2022topological, li2020higher} since we suppose no NNN coupling in our model. For the state in Fig. \ref{fig2}(e)-A whose energy is shown by the green points in Fig.\ref{fig2}(d), its origin is quite counterintuitive since when $t_0=1$ its energy will converge to zero, the energy of singles. As we have shown in Fig. \ref{fig1}(c), there are two singles (e.g., A kind rod at position 4 and B kind rod at position 5 at both edges) near the corner, which is because of the complex structure of 2D-TM systems. Since the coupling $t_1=1-t_0$ between two kinds of rods is not zero now, these two singles can couple to each other and form a new kind of dimers which is named dimers-III. From the field distribution in Fig. \ref{fig2}(e)-C, we can see that the new type of CSs is from the coupling between two dimers-III near the corner. Hence, when $t_0 \rightarrow 1$ ($t_1 \rightarrow 0$), the energy of the CS converges to zero since these coupled singles are almost decoupled from each other. Obviously, the origin of those new two types of CSs are from the complex 2D-TM structure, and we mark them as new types of CSs.

\textit{Photonic HOTIs.}---In this section, we will study the real photonic TM-systems to show that all those HOT states can be realized in real PQCs by strict numerical results from the software without any approximation, i.e., finite-element method (FEM) software COMSOL Multiphysics. Furthermore, from the analyses with defects or randomness, it's found that the new CSs in PQCs could be more robust than the type-II CSs in PhCs. Here, we hope to note that, according to the analysis based on TBM, the coupling terms correspond to the couplings between the rods, not the on-site energies correspond to the Mie resonant frequencies of rods, are the dominant reason for the existing of HOTIs. Hence, in real PQCs, rods-A and rods-B are set to be with different permittivities but the same radius to ensure the relative strengths of couplings between the rods are similar as coupling terms of TBM, but with the side-effect of the different frequencies of Mie resonances\cite{tbm}.

Now, we consider a 2D TM-PQC with $\varepsilon_A=10$, $\varepsilon_B=16$, $\mu_A=\mu_B=1$, $N =4$, and the perfect electrical conductor (PEC) boundary is used, where the distance between PEC and PQC is $d=0.25a$. In Fig. \ref{fig3}, the eigenstates of $E_z$ polarization are shown, in which bulk states, edge states, and corner states at dimers or singles are marked with black, blue, green, and red dots, respectively. We select eight typical states from low frequency to high frequency which are marked with A-H, and $E_z$ field distributions of those states are also shown, where black (grey) circles are rods-A(B). It's easy to find the ESs in Fig. \ref{fig3}-B, -D, and -H. Furthermore, there are three types of CSs in the TM-PQC: non-trivial CS at single rods is shown in Fig. \ref{fig3}-F, CSs at dimers-II are shown in Fig. \ref{fig3}-A and -E, and CSs at dimers-III are shown in Fig. \ref{fig3}-C and -G, and the counterparts of those three types of CSs in TBM can be found in Fig. \ref{fig2}(e)-D, -A, and -C, respectively. The symmetric features of CSs at dimers are also the same as the results of TBM, i.e., CSs of lower frequencies are symmetric along the center of dimers, while CSs of higher frequencies are antisymmetric.

What's more, we can calculate the topological invariant of PQCs which from bulk states by using real space method \cite{SI}. In particular, the dipole moments $p_i^n$ (where $n$ is the n-th gap and $i$ is direction) for the PQC of the first and second gap are $p_x^1=p_y^1=0.5$ and $p_x^2=p_y^2=0$, respectively, and the quadrupole moments of the second gap are near $q_{xy}=0.5$. Proving that the non-trivial type-I CSs at the 1st gap are induced by dipole moments \cite{xie2018second} (shown in Supplemental Material S2), while at the 2nd gap are induced by quadrupole moments \cite{he2020quadrupole,xiong2022topological}. The type-I CSs from both origins can exist in TM-PQC, which are confirmed by a TM-PQC in Supplemental Material S2.
Surprisingly, the solution numbers of type-I CSs of PQCs do not satisfy the filling anomaly theory for traditional type-I CSs of PhCs. The violation maybe from the much more complex photonic band-gap structure of TM-PQC and we will investigate it in further works.

It should be emphasized that the new type of CSs are independent on NNN coupling, so that the new types of CSs found in this work could be more robust than type-II CSs in PhCs, which is the important property for real applications. For example, in Supplemental Material S3, we introduce a defect or randomness to demonstrate this property.

In addition to the results of $S_4$ TM sequence, in Supplemental Material S5 the results of $S_6$ TM sequence are also shown, where CSs are similar to those of $S_4$ TM sequence, except that CSs from dimers are more localized at the corners and easier to be observed.

\textit{Conclusion.}---In summary, we have demonstrated that the topological nontrivial CSs from multipole moments, can be realized in PQCs, whose structure is without translational symmetry. From our limiting knowledge, it's the first time that HOT states are found in PQCs. Even more, new types of CSs are found since the complex structure of PQCs near the corners. The origins of CSs are analyzed by TBM. Our results reveal rich topological physics in PQCs. New CSs extend our understanding of HOT phases and can be used to design novel devices since the all-dielectric structure of PQCs. We also believe CSs widely exist for other waves, e.g., in electronic and phononic QCs, and the research on these topics could be attractive.


\begin{acknowledgments}
	This work is supported by National High Technology Research and Development Program of China (17-H863-04-ZT-001-035-01); National Key Research and Development Program of China (2016YFA0301103, 2018YFA0306201); National Natural Science Foundation of China (12174073). We thank the topology course team for their open-source codes about real space method \cite{RSM}.
\end{acknowledgments}

\bibliographystyle{apsrev4-2}
%


\begin{thebibliography}{38}
\bibitem [{tbm()}]{tbm}%
  \BibitemOpen
  \href@noop {} {\bibinfo {title} {The mie resonant frequencies of one rod in
  real $\rm{PQC}$ can analogy with the on-site energies of one site in
  tight-binding model, and the coupling strength between different rods in real
  $\rm{PQC}$ can analogy with the coupling terms in tight-binding model.
  $\rm{I}$f the permittivities of rods in $\rm{PQC}$ are changed, the mie
  resonant frequency and coupling also will be changed. $\rm{F}$rom the
  analyses of tight-binding model, we can find that the relative magnitudes of
  hopping terms are dominant for the existing of different hoti. $\rm{H}$ence,
  we can change the permittivities of rods to turn the coupling strength, and
  influences of the changes of on-site energies can be almost
  neglected.}}\BibitemShut {Stop}%
\end{thebibliography}
\end{document}